\documentclass[twocolumn,showpacs,preprintnumbers,amsmath,amssymb]{revtex4}

\usepackage{graphicx}
\usepackage{dcolumn}
\usepackage{bm}
\begin{document}

\title{Characteristic frequency of the magnetic radiation
of spinor condensates}

\author{Zhibing Li and Chengguang Bao\footnote{corresponding author: stsbcg@zsu.edu.cn}}

\address{The State Key Laboratory of Optoelectronic Materials and
Technologies \\ School of Physics and Engineering \\ Sun Yat-Sen
University, Guangzhou, 510275, P.R. China}

\begin{abstract}  The magnetic radiation of the
fully-condensed states of $^{23}$Na condensates have been
studied. \ A narrow characteristic spectral line  with a wave length $%
\propto N^{\;-2/5}$ ($N$ is the number of particles) and with  a
probability of transition $\propto N^{17/5}$  emitted (absorbed)
by the condensate was found. \ \ It implies that short wave
radiation with a huge probability of transition can be obtained if
numerous atoms are trapped.  A new technique developed by the
authors, namely, the analytical forms of the fractional parentage
coefficients, was used to calculate analytically the matrix
elements between the total spin-states.
\end{abstract}


\maketitle

The theoretical prediction and the subsequent experimental
realization of scalar and spinor Bose-Einstein condensation (BEC)
\cite{Eins25,Ande95, Davi95,Brad95,Stam98} are great achievements
of scientific research. \ It implies that, under certain
circumstances, BEC as a new kind of aggregations of matter do
exist. One can not rule out the possibility that they might exist
in nature. \ \ If they do exist, an essential way to find them out
is to measure the radiation emitted from them. \ \ \ For this
purpose, one must know the character of the radiation. \ To the
knowledge of the authors, this is a topic not yet been studied
before. \ This paper is an attempt dedicated to the topic. \

For the scalar BEC of neutral atoms, neither charge nor magnetic
moment are contained, therefore no radiation is involved. \
However, the spinor BEC contains magnetic moments, therefore
magnetic multipole radiations may exist. \ For a single atom, the
magnetic moment is small and the related radiation is weak. \
However, for the spinor BEC, the magnetic transition might be
benefited by the Bose-enhancement, because numerous atoms are
involved. \ Thus, an evaluation of the probability and wave length
of the\ transition is meaningful. It would be interesting to see
how large the particle number $N$ is needed so that the magnetic
radiation can be observed.  This is an aim of this paper. For this
purpose, the related initial and final eigenstates together with
their eigenenergies must be known.

 A crucial point is the
calculation of the matix elements for the transition between the
total spin-states each contains numerous particles. When $N$ is
small, this can be achieved by using the fractional parentage
coefficients (FPC) proposed about seventy years ago by Bacher and
Goudsmit.\cite{Bach34} These coefficients are usually obtained
from recursion relations. \cite{Sun89}\ However, when $N$ is large
(say, $N>50$), the procedure of recursion would become extremely
complicated. Therefore the FPC can not be used for many-body
systems previously. \ Recently, we have succeeded to obtain the
analytical forms of these coefficients for spin-1 systems, thereby
they can be used for many-body systems as well disregarding how
large $N$ is \cite{Bao05, Li06,Bao06}. Thus, the main obstacle,
namely, the difficulty in the calculation of related matrix
elements, has been removed.

Consider a condensate of $N$ neutral spin-1 atoms trapped by
parabolic and isotropic confinement so that both the orbital angular
momentum $L$,\  total spin $S$,\  together with their Z-components $M$\ and $%
M_{S}$ are good quantum numbers. Let $\Psi _{L^{\prime }M^{\prime
},S^{\prime }M_{S}^{\prime }}^{f}$ and $\Psi _{LM,S\;M_{S}}^{o}$ be
the final and initial eigenstates,respectively. \ In general, the
rank$-J$ magnetic multipole transition probability per unit time
reads\cite{Rose55}
\begin{equation}
\emph{P}_{M^{\prime },M_{S}^{\prime },\;M,M_{S}}^{(JK)\;f,o}=\beta
^{(J)}|\emph{A}_{M^{\prime },M_{S}^{\prime
},\;M,M_{S}}^{(JK)\;f,o}|^{2}
\end{equation}
\begin{equation}
\beta ^{(J)}=\frac{8\pi (J+1)\;(\omega /c)^{2J+1}\mu _{atom}^{2}}{%
J\lbrack (2J+1)!!\rbrack ^{2}\;\hbar }
\end{equation}
where $\omega $ is the frequency of the emitted (absorbed) photon,
$\mu _{atom}$ is the magnetic moment of the atom.
\begin{equation}
\emph{A}_{M^{\prime },M_{S}^{\prime },\;M,M_{S}}^{(JK)\;f,o}=\langle
\Psi _{L^{\prime }M^{\prime },S^{\prime }M_{S}^{\prime
}}^{f}|\sum^{N}_{j=1} \Xi _{JK,\;j\;}|\Psi
_{LM,\;S\;M_{S}}^{o}\rangle \label{Afo}
\end{equation}
\begin{equation}
\Xi _{JK,\;j\;}=\mathbf{F}_{j}\cdot \nabla
_{j}(r_{j}^{J}Y_{JK}^{\ast }(\stackrel{\wedge }{\mathbf{r}}_{j}))
\end{equation}
where $j$ denotes the $j-th$ atom, $\mathbf{F}_{j}$\ the spin
operator, $Y_{JK}$\ the spherical harmonic, $K$ ranges from $-J$
to $J$. \

Let the components of $\mathbf{F}_{j}$\ be $F_{jx}$, $F_{jy}$, and $%
F_{jz}$, and $F_{j,\pm 1}=\frac{1}{\sqrt{2}}(F_{jx}\pm iF_{jy})$ , $%
F_{j,0}=F_{jz}$. \ For the dipole transition (M1), $\ \Xi _{1,\pm 1,\;j}=\mp
\sqrt{\frac{3}{4\pi }}F_{j,\;\mp 1}$, \ $\Xi _{1,0,\;j}=\sqrt{\frac{3}{4\pi }%
}F_{j,0}$. \ \ For the quadruple transition (M2),$\ \ \Xi _{2,\pm 2,\;j}=\pm
\sqrt{10}r_{j}Y_{1,\mp 1}F_{j,\;\mp 1}$, \ $\ \Xi _{2,\pm 1,\;j}=\mp \sqrt{5}%
r_{j}(Y_{10}F_{j,\;\mp 1}\;\pm Y_{1,\mp 1}F_{j,0})$, \ $\ \Xi _{2,0,\;j}=%
\sqrt{\frac{5}{3}}r_{j}(Y_{11}F_{j,-1}\;-Y_{1,-1}F_{j,+1}+2Y_{1,0}F_{j,0}).$%
\ \ For $J\geq 3$, the transition probabilities would be much smaller and
therefore omitted.

Obviously, the M1 operator keeps all the good quantum numbers
completely unchanged, except $M_{S}$. \ If spin-orbital coupling is
not considered, and if there is no external magnetic field, the
M1-transition can be neglected. \ Therefore, \ we shall focus on the
M2-transition. \
During the transition, $|L^{\prime }-L|\leq 1$, $|S^{\prime }-S|\leq 1$ and $%
M^{\prime }+M_{S}^{\prime }+K=M+M_{S}$ are required. \ In particular, \ the
parity must be changed,

Let $m$ be the mass of the boson \ and $w_{o}$ be the
frequency of the parabolic confinement. \ When $\hbar w_{o}$ and $\sqrt{%
\hbar /mw_{o}}$ are used as units, the Hamiltonian reads
\begin{equation}
H=\frac{1}{2}\sum_{j}(-\nabla _{j}^{2}+r_{j}^{2})+\sum_{i<j}U_{ij}
\end{equation}
where $U_{ij}=\delta (\mathbf{r}_{i}\mathbf{-r}_{j})O_{ij},%
\;O_{ij}=(c_{0}+c_{2}\mathbf{F}_{i}\cdot \mathbf{F}_{j})$. \ \ In
this paper we consider only the transitions related to the fully
condensed (FC) states
\begin{equation}
\Psi _{SM_{S}}^{g}=\Pi _{i=1}^{N}\phi _{S}(r_{i})\vartheta
_{SM_{S}}^{\lbrack N\rbrack } \label{Psig}
\end{equation}
where $\vartheta _{SM_{S}}^{\lbrack N\rbrack }$ is an all-symmetric
normalized total spin-states with the good quantum numbers $S$ and $M_{S}$. $%
S$ is allowed to range from $N,\;N-2,\;N-4,\cdot \cdot \cdot ,$ to 1
or 0.\cite{Bao04,Katr01} The single particle state\ $\phi
_{S}(r)\equiv u_{S}(r)/\sqrt{4\pi }r $ has orbital angular momentum
zero, and it depends on $S$ in general.The spatial wavefunction
$\phi _{S}(r)$ can be obtained by solving the generalized
Gross-Pitaevskii equation\cite{Bao04}
\begin{equation}
\lbrack h_{0}+(N-1)g_{S}\frac{|u_{S}|^{2}}{4\pi r^{2}}%
\;\rbrack u_{S}(r)=\varepsilon _{S}\;u_{S}(r)
\end{equation}
where $h_{0}=\frac{1}{2}\lbrack -\frac{d^{2}}{dr^{2}}+r^{2}\rbrack
$,
\begin{equation}
g_{S}=\langle \vartheta _{SM_{S}}^{\lbrack N\rbrack
}|O_{ij}|\vartheta _{SM_{S}}^{\lbrack N\rbrack }\rangle =c_{0}+c_{2}\frac{%
S(S+1)-2N}{N\;(N-1)}
\end{equation}
and $\varepsilon _{S}$ is the chemical potential. The eigenenergy of
a FC-state depends surely on $S$, it reads
\begin{equation}
E_{S}^{g}=N\stackrel{\_}{h}_{0}+\frac{N\;(N-1)}{2}%
g_{S}T_{S} \label{Eg}
\end{equation}
where $\stackrel{\_}{h}_{0}=\int dr\;u_{S}h_{0}u_{S}$ and $T_{S}=\frac{1}{%
4\pi }\int_{0}^{\infty }\frac{dr}{r^{2}}|u_{S}|^{4}.$ Since $E_{S}^{g}$
depends on $S$, all the FC-states with $S$ ranged from $N,$ to 1 or 0 spread
into  a band, the FC-band.\cite{Bao04}

It is obvious that the M2-operators can excite only one $p-$wave
particle. \ In other words, a FC-state would transit to a final
state with one particle excited. From the rule of outer product
$\{1\}\otimes \{N-1\}=\{N\}+\{N-1,1\}$, the spatial wave functions
of the final states may have two choices of permutation symmetry
$\lambda $. The choice $\lambda =\{N\}$ is just an excitation of the
center of mass (c.m.), the related photon energy is just $\hbar
w_{o}$. \ \ This is a trivial case with a very low photon energy. \
For the choice $\lambda =\{N-1,1\},$ the final state with an excited
particle can be written as \cite{Bao05}
\begin{equation}
\Psi _{M,SM_{S}}^{ex}=\frac{1}{\sqrt{N}}\lbrack
\sum_{i=1}^{N}\;\varphi _{M,S}^{b}\mathbf{(r}_{i})\; \Pi _{j \neq
i}\varphi _{S}^{a}(r_{j})\;\Theta _{SM_{S}}^{\lbrack N\rbrack
,i}\rbrack \label{Psiex}
\end{equation}
where $N-1$ particles condense into the same $\varphi _{S}^{a}(r)=\frac{1}{%
\sqrt{4\pi }\;r}u_{S}^{a}(r)$ , while a particle is excited to a higher
state $\varphi _{M,S}^{b}(\mathbf{r})=\frac{1}{\;r}u_{S}^{b}(r)Y_{1M}(%
\stackrel{\wedge }{\mathbf{r}})$\ , $\Theta _{SM_{S}}^{\lbrack N\rbrack ,i}$
is a normalized total spin-state with the good quantum numbers $S$ and $%
M_{S} $, and with $\lambda =\{N-1,1\}$, where $S$ is allowed to
range from 1 to $N-1$. \cite{Bao05, Katr01} \ Obviously, the
summation over $i$ \ in (10) assures that $\Psi _{M,SM_{S}}^{ex}$
is all-symmetric as required.

 $u_{S}^{a}$ and $u_{S}^{b}$ satisfy
the set of coupled equations \cite{Bao05}
\begin{equation}
\lbrack h_{0}+g_{S}^{b}\frac{|u_{S}^{b}|^{2}}{4\pi
r^{2}}+(N-2)g_{S}^{a}\frac{|u_{S}^{a}|^{2}}{4\pi r^{2}}\rbrack
\;u_{S}^{a}=\varepsilon _{a}u_{S}^{a} \label{h0}
\end{equation}
\begin{equation}
\lbrack h_{1}+(N-1)g_{S}^{b}\frac{%
|u_{S}^{a}|^{2}}{4\pi r^{2}}\rbrack \;u_{S}^{b}=\varepsilon
_{b}u_{S}^{b} \label{h1}
\end{equation}
where $h_{1}=\frac{1}{2}\lbrack -\frac{d^{2}}{dr^{2}}+\frac{2}{r^{2}}%
+r^{2}\rbrack ,$\ $g_{S}^{a}=\langle \Theta _{SM_{S}}^{\lbrack N\rbrack
,i}|O_{jk}|\Theta _{SM_{S}}^{\lbrack N\rbrack ,i}\rangle $ ( $i\neq j\neq k$%
), and $\ g_{S}^{b}=\langle \Theta _{SM_{S}}^{\lbrack N\rbrack
,i}|O_{ij}|\Theta _{SM_{S}}^{\lbrack N\rbrack ,i}+\Theta
_{SM_{S}}^{\lbrack N\rbrack ,j}\rangle $. \ By using the FPC,
$g_{S}^{a}$ and $g_{S}^{b}$ can be derived as\cite{Bao05}
\begin{eqnarray}
g_{S}^{a}&=&c_{0}+ c_{2}\frac{1\;}{(N-1)(N-2)} \nonumber \\ & & \lbrack \frac{%
N+1+(-1)^{N-S}}{N}S(S+1)-2(N-1)\rbrack
\end{eqnarray}
\begin{equation}
g_{S}^{b}=\frac{N-2}{N-1}(c_{0}+c_{2})-(1+(-1)^{N-S})\frac{S(S+1)}{%
N\;(N-1)}c_{2}
\end{equation}

Since $g_{S}^{a}$ and $g_{S}^{b}$ are known, eigenstates can be obtained by
solving (11) and (12),\ the associated eigenenergy reads
\begin{eqnarray}
E_{S}^{ex}&=&\stackrel{\_}{h}_{1}^{b}+(N-1)\stackrel{\_}{%
h}_{0}^{a}+(N-1)g_{S}^{b}T_{S}^{ab} \nonumber
\\
& &+\frac{(N-1)(N-2)}{2}g_{S}^{a}T_{S}^{a} \label{Eex}
\end{eqnarray}
where $\stackrel{\_}{h}_{1}^{b}=\int dr\;u_{S}^{b}h_{1}u_{S}^{b}$ , $%
\stackrel{\_}{h}_{0}^{a}=\int dr\;u_{S}^{a}h_{0}u_{S}^{a}$\ , $T_{S}^{ab}=%
\frac{1}{4\pi }\int_{0}^{\infty }\frac{dr}{r^{2}}|u_{S}^{a}|^{2}$\ $%
|u_{S}^{b}|^{2}$, and $T_{S}^{a}=\frac{1}{4\pi }\int_{0}^{\infty }\frac{dr}{%
r^{2}}|u_{S}^{a}|^{4}.$

From now on let $S$ ($S^{\prime }$) denotes the total spin of the
initial (final) state.  For M2 transitions we have $S^{\prime }=S$
or $S\pm 1$. From (15) and (9), the energy difference $(\Delta
E)_{S^{\prime }S}=E_{S^{\prime }}^{ex}-E_{S}^{g}$ can be known (
$|(\Delta E)_{S^{\prime }S}|$\ is the photon energy).

By solving eq.(7), (11), and (12), not only the photon energies
can be known, the probability of M2-transition can also be
obtained. \ To compare with experimental observation, we define
\begin{equation}
\emph{P}^{\;f,o}_{S^{\prime }S}=\frac{1}{2S+1}\sum_{M^{\prime
},M_{S}^{\prime },M_{S}}\emph{P}_{M^{\prime }M_{S}^{\prime
},\;MM_{S}}^{(2K)\;f,o}
\end{equation}

By inserting (6) and (10) into (3), we have
\begin{equation}
\emph{P}^{\;f,o}_{S^{\prime }S}=\frac{50%
}{12\pi }\frac{N}{2S+1}\beta ^{(2)}(I_{S^{\prime
}S})^{2}Y_{S^{\prime }S} \label{Pfo}
\end{equation}
where $I_{S^{\prime }S}=\int dr\;u_{S^{\prime }}^{b}ru_{S}$ and
\begin{eqnarray}
Y_{S^{\prime }S}&=&\sum_{M_{S}}\{|\langle \Theta _{S^{\prime
},M_{S}+1}^{\lbrack N\rbrack ,1}|F_{1,1}|\vartheta
_{SM_{S}}^{\lbrack N\rbrack }\rangle |^{2}\nonumber \\
& &+|\langle \Theta _{S^{\prime },M_{S}}^{\lbrack N\rbrack
,1}|F_{1,0}|\vartheta _{SM_{S}}^{\lbrack N\rbrack }\rangle |^{2}
\nonumber \\
& &+|\langle \Theta _{S^{\prime },M_{S}-1}^{\lbrack N\rbrack
,1}|F_{1,-1}|\vartheta _{SM_{S}}^{\lbrack N\rbrack }\rangle |^{2}\}
\label{Yss}
\end{eqnarray}

It turns out that the quantity inside the brackets in (18) does
not depend on $M_{S}$. \ By using the FPC and refer to the
appendix of \cite{Li06}, we have
\begin{equation}
Y_{S^{\prime }S}=(2S+1)(N-S)(N+S+1)/N^{2}
\end{equation}
\begin{equation}
Y_{S^{\prime }S}=(N-S)(S+2)/N
\end{equation}
 and
\begin{equation}
Y_{S^{\prime }S}=(N+S+1)(S-1)/N
\end{equation}
if\; $S^{\prime }=S, S+1, and S-1$, respectively.

The numerical results of Na atoms are presented in the follows as
examples. The
interaction of Na has $c_{0}=6.774\times 10^{-4}\sqrt{w_{o}}$ and $%
c_{2}=2.117\times 10^{-5}\sqrt{w_{o}}$, $w_{o}=300\sec ^{-1}$ is
given in this paper. $(\Delta E)_{S^{\prime }S}$ are given in
Fig.1. \ There are three points noticeable. (i) For the $S$ to
$S'=S$ transition, there is a turning point appearing at
$S/N=1/\sqrt{3}$\ . This is associated with a sudden change of the
structure of the excited particle as shown later.\ (ii) $(\Delta
E)_{S^{\prime }S}$ can be either positive or negative. \ It
implies that, when $N$ is very large, the FC-states may not be the
lowest. (iii)  The magnitude of $(\Delta E)_{S^{\prime }S}$ of the
$S$ to $S'=S-1$ transition is relatively small, thus the
probability of this transition is negligible.

The probability $\emph{P}^{\;f,o}_{S^{\prime }S}$ is given in
Fig.2, where only the curves for $S'=S$ transition (solid line)
and for $S'=S+1$ transition (dashed line) can be seen. \ There is
a striking sharp peak at $S/N=1/\sqrt{3}$ for the $S'=S$
transition (just at the turning point of $(\Delta E)_{S^{\prime
}S}$ ), the height of the peak is $1.44\times 10^{-34}/sec$, it is
so high that only the bottom of the peak was seen in the figure.\
Accordingly, there would be a narrow spectral line emerged more
than ten times brighter than the one contributed by the lower
peak. \ The lower peak is much lower and much broader, it is
peaked at $S/N=0.96$ and is contributed by the $S'=S+1$
transitions. From Fig.1 we know that the narrow peak has photon
energy 127.9, while the broad peak has photon energies ranging
from about 220 to 380. Thus, the feature of the spectrum is clear,
namely, a bright narrow line together with a background at its
violet side. The narrow line is called the characteristic line of
the system thereafter.

When $N$ varies, the features of Fig.1 and 2 remain unchanged, but
the magnitudes of  $(\Delta E)_{S^{\prime }S}$ and
$\emph{P}^{\;f,o}_{S^{\prime }S}$ are changed remarkably. In Fig.3
the wave lengths  $\lambda _{S^{\prime }S}$ associated with the
photon energies of the narrow peak and the lower peak against
$log(N)$ are plotted.  Where the two parallel straight lines imply
that $\lambda _{S^{\prime }S}$ are proportional to $N^{-2/5}$ in
both cases, thus very short wave radiation can be obtained if $N$ is
sufficiently large. In Fig.4 the probabilities $P_{S^{\prime
}S}^{f,o}$ are plotted. The solid line as a straight line implies
that the probability of transition of the characteristic line is
proportional to $N^{17/5}$. \ The power 17/5 is striking, it implies
that the probability can be terribly large when $N$\ is very large.
The dashed line implies that the probability associated with the
$S'=S+1$ transition is much lower, and it increases with $N$ much
slower. \ For a quantitative example, if $N>10^{24}$,
 $P_{S^{\prime }S}^{f,o}$ of the characteristic line would be
$>10^{4}\sec ^{-1}$\ and $\lambda _{S^{\prime }S}<150cm$.

Let us study the physical background of the above findings. When
$N$ is very large, we can neglected the second term of (11), then
$u_{S^{\prime }}^{a}$ can be obtained by using further the
Thomas-Fermi approximation,\cite{Baym96} (neglecting the kinetic
energy).  When the $u_{S^{\prime }}^{a}$ obtained in this way is
inserted into (12), the equation can be rewritten as
\begin{equation}
\lbrack \frac{1}{2}(-\frac{d^{2}}{dr^{2}}+\frac{2}{r^{2}}%
)+U_{eff}\rbrack u_{S^{\prime }}^{b}=\varepsilon _{b}u_{S^{\prime
}}^{b} \label{TFA1}
\end{equation}
Where
\begin{equation}
U_{eff}= \frac{g_{S^{\prime }}^{b}}{2 g_{S^{\prime
}}^{a}}(r_{a}^{2}-r^{2})\zeta (r_{a}-r)+\frac{r^{2}}{2} \label{TFA2}
\end{equation}
where $r_{a}=(15Ng_{S^{\prime }}^{a}/4\pi )^{1/5}$ and  $\zeta
(r_{a}-r)=1$ if $r\leq r_{a}$\ , or $=0$ otherwise.

It turns out that the feature of $U_{eff}$ is extremely sensitive
to $g_{S^{\prime }}^{b}/g_{S^{\prime }}^{a}$.   \ When $N-S'$ is
even and $S'$ is close to $N/\sqrt{3}$ (or $N-S'$ is odd and $S'$
is close to $N$),  $g_{S^{\prime
}}^{b}/g_{S^{\prime }}^{a}\approx 1,$ thereby $U_{eff}$ is nearly a constant in the broad domain  $%
r<r_{a}$, but has a sharp border at $r_{a}$ as shown by the
$S'=/\sqrt{3}$ line of Fig.5. In this special case, $u_{S^{\prime
}}^{b}$\ spreads widely from 0 to $r_{a}$, and results in having a very large $%
I_{S^{\prime }S}$, and thereby a very large probability of transition. Alternatively, when $%
g_{S^{\prime }}^{b}/g_{S^{\prime }}^{a}<1, U_{eff}$ is attractive
as shown by the $S'=N$ curve of Fig.5.  In this case the excited
particle
 would be distributed close to the center.  When $%
g_{S^{\prime }}^{b}/g_{S^{\prime }}^{a}>1, U_{eff}$ is repulsive
as shown by the $S'=0$ curve of Fig.5.  In this case the excited
particle
 would be distributed very close to the surface of the condensate.
  To give quantitative data related to Fig.5, $u_{S^{\prime
}}^{b}$ is mainly distributed in the domain $0<r<11$ if $S'=N$,
$0<r<148.5$ if $S'=N/\sqrt{3} $, and $145.5<r<148.5$ if $S'=0$.
These data exhibit the great difference in the wave functions of
the excited particle, which results in the dramatic variation of
$I_{S^{\prime }S}$. When $S'=S$ we have $I_{S^{\prime }S}=4.24$,
14.07, 87.21, 0.81, and $0.04$ if $S'=0$, $57N/100$, $N/\sqrt{3}$,
$58N/100$, and $N$, respectively. These data exhibit that, when
$S'$\ varies across $N/\sqrt{3}$, a great increase and a great
decrease of $I_{S^{\prime }S}$ occur successively, this is the
origin of the narrow peak shown in Fig.2.

For the transition with $N-S'$ odd and $S'=S+1$, a similar
increase of $I_{S^{\prime }S}$ occurs if $S'$ is approaching $N$.
However, in this case $Y_{S^{\prime }S}$ tends to zero (refer to
eq.(20)). As a result of the competition of  $I_{S^{\prime }S}$
and $Y_{S^{\prime }S}$, the lower and broad peak appears adjacent
to $S=N$ as shown in Fig.2.

In summary, we have found that the spin-dependent interaction is
crucial to the magnetic radiation of spinor BEC.  The effect of the
interaction is embodied by the ratio $g_{S^{\prime
}}^{b}/g_{S^{\prime }}^{a}$, which affects $U_{eff}$ seriously.
Accordingly, the excited particle has three choices of structure,
namely, being close to the center, close to the outer surface, and
widely distributed from the center to the surface.  The wide
distribution leads to a great increase of
$\emph{P}^{\;f,o}_{S^{\prime} S}$.  However, this case occurs only
if $S'\approx N/\sqrt{3} $ and $S'=S$, or $S'\approx N $ and
$S'=S+1$. This leads to the appearance of the characteristic line
together with a broad background as shown in Fig.2.  The
N-dependence of $\lambda _{S^{\prime }S}$ and
$\emph{P}^{\;f,o}_{S^{\prime} S}$ has been studied, and a $N^{-2/5}$
and $N^{17/5}$ dependence, respectively, was found. This implies
that short-wave radiation with a large probability of transition can
be achieved if numerous particles are trapped.

It is mentioned that only the magnetic radiation of the FC-states
has been studied in this paper, it is still a long way to
understand the spectroscopy of the spinor BEC.

Acknowledgment: We appreciate the support from the NSFC under the
grants 10574163, 90306016, and 10674182.

\clearpage

\begin{figure}
\includegraphics{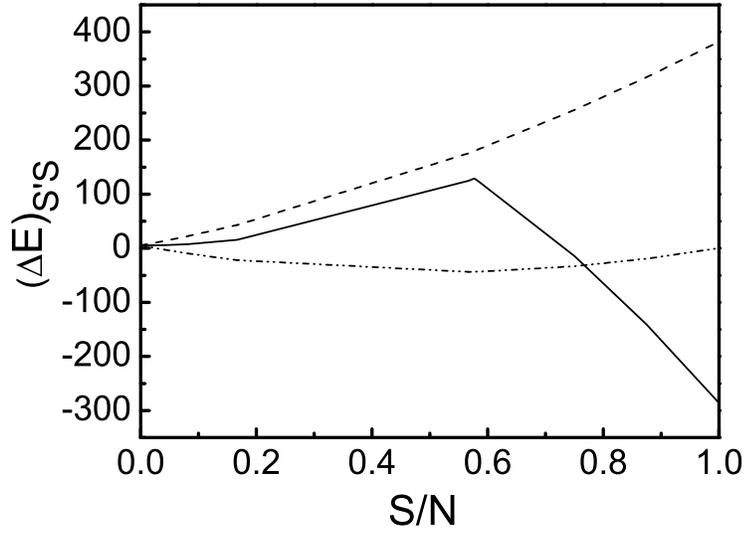}
\caption{\label{fig:1} $(\Delta E)_{S^{\prime }S}=E_{S^{\prime }}^{ex}-E_{S}^{g}$ \ against $%
S/N$ for $^{23}$Na atoms with\ $N=5\times 10^{12}.$ The solid
curve is for $S^{\prime }=S$ , the dashed curve is for $S^{\prime
}=S+1$, the dash-dot-dot curve is for $S^{\prime }=S-1$.
$w_{o}=300\sec ^{-1}$, the unit of energy is $\hbar w_{o}$.  The
notations and the value of $w_{o}$ are the same in the following
figures. }
\end{figure}

\begin{figure}
\includegraphics{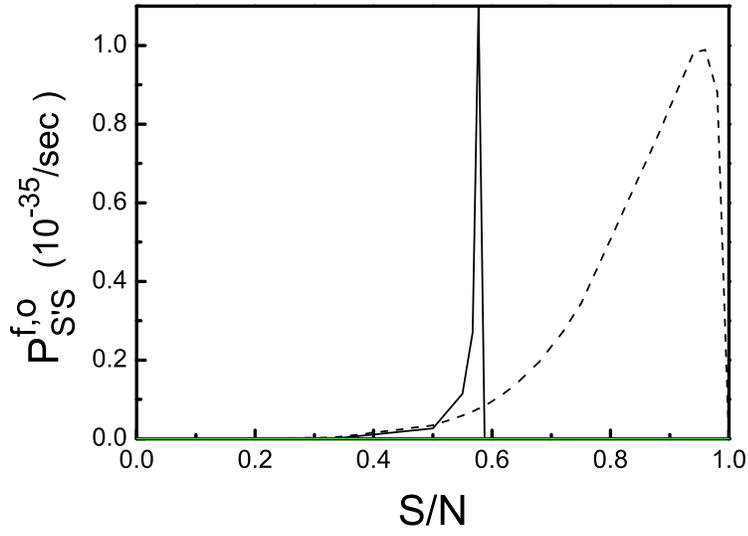}
\caption{\label{fig:2} The probability $\emph{P}_{S^{\prime
}S}^{\;f,o}$ of the M2-transition of the FC-states of $^{23}$Na
against $S/N$. $N=5\times 10^{12}$ . }
\end{figure}

\begin{figure}
\includegraphics{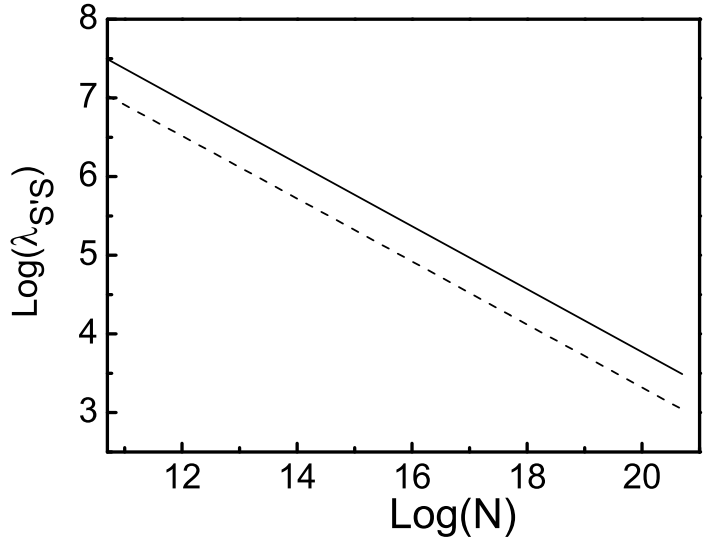}
\caption{\label{fig:3} The wave lengths $\lambda _{S^{\prime }S}$
(in cm) of the radiation of the FC-states against $log(N)$.
 The solid line is for the characteristic line which has $S=N/\sqrt{3} $ and $S'=S$, the dashed
line is for the lower peak which has $S=96N/100$ and $S'=S+1$. }
\end{figure}

\begin{figure}
\includegraphics{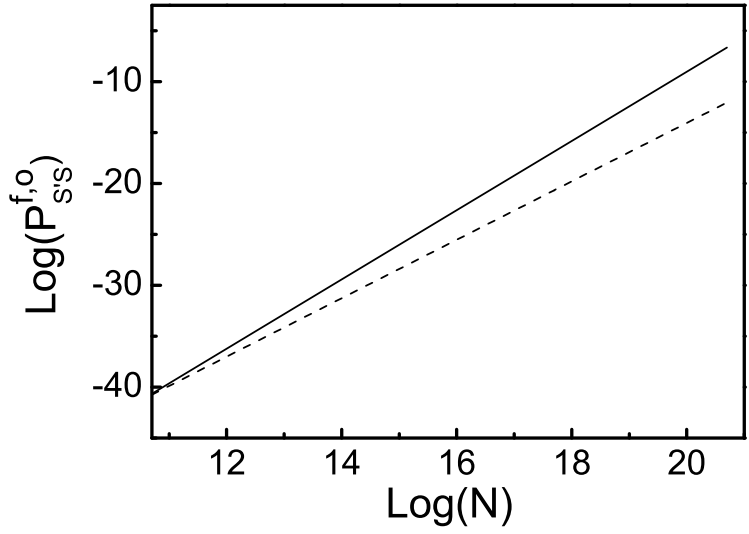}
\caption{\label{fig:4} The probability of transition (in $%
\sec ^{-1}$) against $log(N)$. Refer to Fig.3 .}
\end{figure}

\begin{figure}
\includegraphics{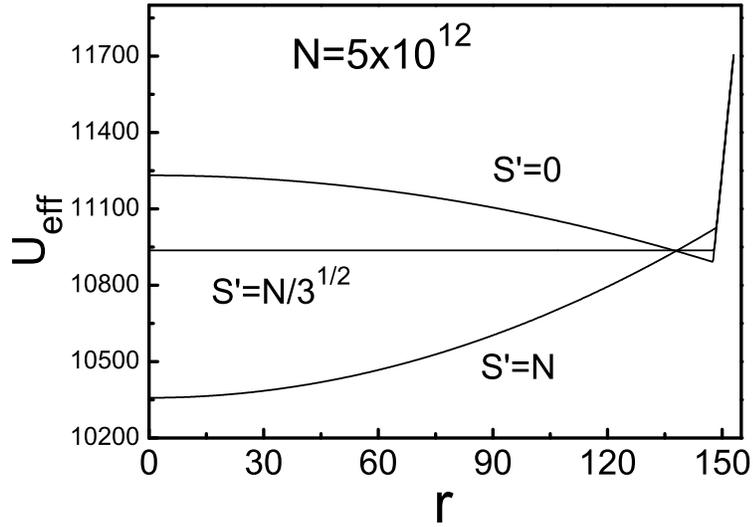}
\caption{\label{fig:5}  $U_{eff}(r)$ for the excited particle with
three cases of $S^{\prime }$, $N-S^{\prime }$ is assumed to be
even.  The units of energy and length are $\hbar w_{o}$ and $\sqrt{%
\hbar /mw_{o}}$, respectively. }
\end{figure}

\end{document}